\documentclass[pra,aps,nopacs,onecolumn,twoside,superscriptaddress]{revtex4}

\usepackage{graphicx}
\usepackage{dcolumn}
\usepackage{bm}

\usepackage{amsmath,amsfonts,amssymb,dsfont,hyperref,color,epsfig,graphics,graphicx,latexsym,mathrsfs,revsymb,theorem,url,verbatim,epstopdf,cleveref,dashrule,cases,booktabs,diagbox,threeparttable,setspace,natbib,mathdots}

\hypersetup{colorlinks,linkcolor={blue},citecolor={blue},urlcolor={red}}

\newtheorem{definition}{Definition}

\newtheorem{lemma}[definition]{Lemma}

\newtheorem{theorem}[definition]{Theorem}

\def\squareforqed{$\square$}
\def\qed{\ifmmode\squareforqed\else{\unskip\nobreak\hfil
\penalty50\hskip1em\null\nobreak\hfil\squareforqed
\parfillskip=0pt\finalhyphendemerits=0\endgraf}\fi}

\def\endenv{\ifmmode\;\else{\unskip\nobreak\hfil
\penalty50\hskip1em\null\nobreak\hfil\;
\parfillskip=0pt\finalhyphendemerits=0\endgraf}\fi}

\newenvironment{proof}{\noindent \textbf{{Proof.~} }}{\qed}

\def\bpf{\begin{proof}}
\def\epf{\end{proof}}
\def\bea{\begin{eqnarray}}
\def\eea{\end{eqnarray}}
\def\beq{\begin{equation}}
\def\eeq{\end{equation}}
\def\bal{\begin{aligned}}
\def\eal{\end{aligned}}
\def\bma{\begin{bmatrix}}
\def\ema{\end{bmatrix}}

\def\dim{\mathop{\rm Dim}}

\def\dg{\dagger}

\def\a{\alpha}

\def\s{\sigma}

\def\ps{\psi}

\newcommand{\ket}[1]{|{#1}\rangle}
\newcommand{\proj}[1]{|{#1}\rangle \langle {#1}|}

\newcommand{\nc}{\newcommand}

\nc{\bbA}{\mathbb{A}} \nc{\bbB}{\mathbb{B}} \nc{\bbC}{\mathbb{C}}
\nc{\bbD}{\mathbb{D}} \nc{\bbE}{\mathbb{E}} \nc{\bbF}{\mathbb{F}}
\nc{\bbG}{\mathbb{G}} \nc{\bbH}{\mathbb{H}} \nc{\bbI}{\mathbb{I}}
\nc{\bbJ}{\mathbb{J}} \nc{\bbK}{\mathbb{K}} \nc{\bbL}{\mathbb{L}}
\nc{\bbM}{\mathbb{M}} \nc{\bbN}{\mathbb{N}} \nc{\bbO}{\mathbb{O}}
\nc{\bbP}{\mathbb{P}} \nc{\bbQ}{\mathbb{Q}} \nc{\bbR}{\mathbb{R}}
\nc{\bbS}{\mathbb{S}} \nc{\bbT}{\mathbb{T}} \nc{\bbU}{\mathbb{U}}
\nc{\bbV}{\mathbb{V}} \nc{\bbW}{\mathbb{W}} \nc{\bbX}{\mathbb{X}}
\nc{\bbY}{\mathbb{Y}} \nc{\bbZ}{\mathbb{Z}}

\nc{\bA}{{\bf A}} \nc{\bB}{{\bf B}} \nc{\bC}{{\bf C}}
\nc{\bD}{{\bf D}} \nc{\bE}{{\bf E}} \nc{\bF}{{\bf F}}
\nc{\bG}{{\bf G}} \nc{\bH}{{\bf H}} \nc{\bI}{{\bf I}}
\nc{\bJ}{{\bf J}} \nc{\bK}{{\bf K}} \nc{\bL}{{\bf L}}
\nc{\bM}{{\bf M}} \nc{\bN}{{\bf N}} \nc{\bO}{{\bf O}}
\nc{\bP}{{\bf P}} \nc{\bQ}{{\bf Q}} \nc{\bR}{{\bf R}}
\nc{\bS}{{\bf S}} \nc{\bT}{{\bf T}} \nc{\bU}{{\bf U}}
\nc{\bV}{{\bf V}} \nc{\bW}{{\bf W}} \nc{\bX}{{\bf X}}
\nc{\bY}{{\bf Y}} \nc{\bZ}{{\bf Z}}

\nc{\bmA}{{\bm A}} \nc{\bmB}{{\bm B}} \nc{\bmC}{{\bm C}}
\nc{\bmD}{{\bm D}} \nc{\bmE}{{\bm E}} \nc{\bmF}{{\bm F}}
\nc{\bmG}{{\bm G}} \nc{\bmH}{{\bm H}} \nc{\bmI}{{\bm I}}
\nc{\bmJ}{{\bm J}} \nc{\bmK}{{\bm K}} \nc{\bmL}{{\bm L}}
\nc{\bmM}{{\bm M}} \nc{\bmN}{{\bm N}} \nc{\bmO}{{\bm O}}
\nc{\bmP}{{\bm P}} \nc{\bmQ}{{\bm Q}} \nc{\bmR}{{\bm R}}
\nc{\bmS}{{\bm S}} \nc{\bmT}{{\bm T}} \nc{\bmU}{{\bm U}}
\nc{\bmV}{{\bm V}} \nc{\bmW}{{\bm W}} \nc{\bmX}{{\bm X}}
\nc{\bmY}{{\bm Y}} \nc{\bmZ}{{\bm Z}}

\nc{\cA}{{\cal A}} \nc{\cB}{{\cal B}} \nc{\cC}{{\cal C}}
\nc{\cD}{{\cal D}} \nc{\cE}{{\cal E}} \nc{\cF}{{\cal F}}
\nc{\cG}{{\cal G}} \nc{\cH}{{\cal H}} \nc{\cI}{{\cal I}}
\nc{\cJ}{{\cal J}} \nc{\cK}{{\cal K}} \nc{\cL}{{\cal L}}
\nc{\cM}{{\cal M}} \nc{\cN}{{\cal N}} \nc{\cO}{{\cal O}}
\nc{\cP}{{\cal P}} \nc{\cQ}{{\cal Q}} \nc{\cR}{{\cal R}}
\nc{\cS}{{\cal S}} \nc{\cT}{{\cal T}} \nc{\cU}{{\cal U}}
\nc{\cV}{{\cal V}} \nc{\cW}{{\cal W}} \nc{\cX}{{\cal X}}
\nc{\cY}{{\cal Y}} \nc{\cZ}{{\cal Z}}

\nc{\hA}{{\hat{A}}} \nc{\hB}{{\hat{B}}} \nc{\hC}{{\hat{C}}}
\nc{\hD}{{\hat{D}}} \nc{\hE}{{\hat{E}}} \nc{\hF}{{\hat{F}}}
\nc{\hG}{{\hat{G}}} \nc{\hH}{{\hat{H}}} \nc{\hI}{{\hat{I}}}
\nc{\hJ}{{\hat{J}}} \nc{\hK}{{\hat{K}}} \nc{\hL}{{\hat{L}}}
\nc{\hM}{{\hat{M}}} \nc{\hN}{{\hat{N}}} \nc{\hO}{{\hat{O}}}
\nc{\hP}{{\hat{P}}} \nc{\hR}{{\hat{R}}} \nc{\hS}{{\hat{S}}}
\nc{\hT}{{\hat{T}}} \nc{\hU}{{\hat{U}}} \nc{\hV}{{\hat{V}}}
\nc{\hW}{{\hat{W}}} \nc{\hX}{{\hat{X}}} \nc{\hZ}{{\hat{Z}}}

\nc{\hn}{{\hat{n}}}

\def\dim{\mathop{\rm Dim}}

\def\dg{\dagger}

\begin{document}


\title{Nonlocality via multiqubit orthogonal product bases}



\author{Lin Chen}
\email[]{linchen@buaa.edu.cn(corresponding author)}
\affiliation{LMIB and School of Mathematical Sciences, Beihang University, Beijing 100191, China}
\affiliation{International Research Institute for Multidisciplinary Science, Beihang University, Beijing 100191, China}

\author{Yutong Jiang}  
\email[]{mitiaojiang@163.com}
\affiliation{Beijing No.4 High School International Campus, Beijing, 100031, China}

\date{\today}

\Large

\begin{abstract}
We investigate the quantum nonlocality via the discrimination on two, three and four-qubit orthogonal product bases (OPBs). We show that every two-qubit, and some three and four-qubit OPBs can be locally distinguished. It turns out that the  remaining three and four-qubit OPBs cannot be locally distinguished, and thus they show the quantum nonlocality without entanglement. We also distinguish them by merging some systems using quantum teleportation with assisted Bell states.
\end{abstract} 
                             
\maketitle


\vspace{1cm}

keywords: orthogonal product basis,  multiqubit, teleportation


\section{INTRODUCTION}
\label{sec:int} 

Quantum entanglement manifests quantum nonlocality, and plays a key role in quantum cryptography, computing and operations \cite{bennett1992communication,barenco1995elementary,paraiso2021photonic,Schaetz2004quantum,chen2005probabilistic,mattle1996dense}. In contrast,
multipartite product states can be prepared under local operations and classical communications (LOCC) and are a kind of free resource in practice. They are closely related to the separability problem \cite{hhh96}, entanglement theory \cite{bds96} and resource theory \cite{faithful2022}. Nevertheless, a set of orthogonal product states such as unextendible product basis (UPB) may show quantum applications including the local indistinguishability \cite{BDF+99,AL01} and strong nonlocality
\cite{quantum2021}, as well as unextendible product operator basis \cite{upob2022}. A special UPB, namely the so-called orthogonal product basis (OPB), spans the whole space containing the OPB. For example, every multiqubit space is spanned by a multiqubit OPB. So it is a physically operational to investigate the multiqubit space in terms of OPBs. 
Technically, one can construct infinitely many OPBs. The investigation on the construction and local distinguishability of some OPBs, as well as the connection to other notions like uncompletable product basis has been carried out in recent years \cite{dms03,Feng2009Characterizing,cdjpa2017,Shi2022}, though a more complete picture is far from satisfaction yet. As far as we know, the local distinguishability of few qubit OPBs is incomplete. 

On the other hand, quantum teleportation is a prominent application in quantum information \cite{bbc93}. 
Plenty of theoretical and experimental study has been devoted to the realization of teleportation in the past decades \cite{yeo2006teleportation,agrawal2006perfect,fiaschi2021optomechanical,fonseca2019highdim,graham2015superdense,harris2021qt,pirandola2017fundamental,ren2017ground,luo2019qt}. Teleportation requires a Bell state as the necessary and sufficient quantum resource under LOCC. Teleportation can gather information of distributed systems at the cost of quantum entanglement. It indicates that teleportation may help distinguish OPBs of distributed systems.

In this paper, we locally distinguish $n$-qubit OPBs when $n=2,3$ and $4$. They have been fully classified in the paper \cite{cdjpa2017}. We investigate them in Theorems \ref{thm:2qubit}, \ref{thm:3qubit} and \ref{thm:4qubit}, respectively. For this purpose, we present the notion of irreducible OPBs and their properties in Lemmas \ref{le:reducible=SUMirreducible} and \ref{le:multiplicity<n}. Such OPBs show the quantum nonlocality without entanglement.  Then we merge two systems of four-qubit OPBs by teleportation at the cost of one ebit, and are able to locally distinguish the resulting $2\times4$ and $2\times2\times4$ OPBs. They are presented in Theorems \ref{thm:3qubit-qt} and \ref{thm:4qubit-qt}, respectively. Our results show that the entanglement cost for local distinguishability of multiqubit OPBs might be constant in spite of the increase of qubit number. 

The rest of this paper is organized as follows. In Sec. \ref{sec:pre} 
we introduce the preliminary definitions and facts. In Sec. \ref{sec:23qubit}, we show the local distinguishability of two-qubit, and some three and four-qubit OPBs. We also present the general condition. In Sec. \ref{sec:4qubit}, we investigate the local distinguishability of three and four-qubit OPBs by teleportation. Finally we conclude in Sec.  
\ref{sec:con}.

\section{PRELIMINARIES}
\label{sec:pre}

We work with the $n$-qubit system in $\bbC^2\otimes...\otimes\bbC^2$, for $n=2,3$ and $4$. We refer to the systems as $A,B,C$ and $D$, respectively. They are practically short for Alice, Bob, Charlie and Daniel, respectively. We also refer to $\ket{a,b}\in\bbC^m\otimes\bbC^n$ as a bipartite product state. One can similarly define a tripartite state $\ket{a,b,c}$ and so on. An orthogonal product basis (OPB) in the $n$-partite space $\bbC^{d_1}\otimes...\otimes\bbC^{d_n}$ is a set of orthogonal product states spanning the space. Evidently, every space has the \textit{trivial} OPB $\{\ket{i_1,...,i_n},\quad i_j=0,1,...,d_j-1, \quad j=1,2,...,n\}$. To efficiently classify OPBs, we say that two sets of $n$-partite states $\cA$ and $\cB$ are equivalent when up to the system permutation $\s$ there is a product unitary matrix $U=U_1\otimes...\otimes U_n$ such that, every vector $\ket{\ps}\in\cA$ satisfies that $U\ket{\ps}_{\s(A_1...A_n)}$ is proportional to a state in $\cB$. Hence,  
every set equivalent to an OPB is still an OPB. Because unitary gates are realizable with certainty, the following observation is clear.
\begin{lemma}
\label{le:OPB=distinguishableATtheSAMEtime}	
Two equivalent OPBs are locally distinguishable at the same time.
\end{lemma}
As a result, we only need to distinguish one OPB $\cA$ out of a family of OPBs equivalent to the $\cA$. In the following, we list a few examples of two and three-qubit OPBs. 
\begin{eqnarray}
\label{eq:M2}
M_2=
\bma
0 & b \\
0 & b' \\
1 & c \\ 
1 & c' 
\ema
:=
\bma
0 & * \\
1 & * 
\ema,
\quad
M_{31}=
\bma
0 & 0 & 0 \\
a & 1 & 0 \\
a' & 1 & 0 \\
0 & b & 1 \\
0 & b' & 1 \\
1 & 0 & c \\
1 & 0 & c' \\
1 & 1 & 1 
\ema
:=
\bma
0 & 0 & 0 \\
* & 1 & 0 \\
0 & * & 1 \\
1 & 0 & * \\
1 & 1 & 1 
\ema,
\end{eqnarray}
and
\begin{eqnarray}
\label{eq:M32}
M_{32}=
\bma
0 & 0 & * \\
0 & 1 & * \\
1 & e & * \\
1 & e' & * \\
\ema,
\quad
M_{33}=
\bma
0 & 0 & * \\
0 & 1 & * \\
1 & * & e \\
1 & * & e' \\
\ema.	
\end{eqnarray}
Here, the star $*$ means shorthand for the $2\times1$ submatrix $\bma b \\ b' \ema$, $\bma c \\ c' \ema$ and so on. In other words, the stars in a matrix represent distinct submatrices. The first row of $M_2$ refers to the product state $\ket{0,b}$, and 
the matrix $M_2$ in \eqref{eq:M2} represents the two-qubit OPB consisting of $\ket{0,b},\ket{0,b'},\ket{1,c}$ and $\ket{1,c'}$. The symbols $b,b'$ refer to a qubit orthonormal basis $\ket{b},\ket{b'}$, and we shall also refer to $b$ as a vector variable for simplicity. For convenience we may refer to $M_2$ as the orthogonal product matrix (OPM) of the two-qubit OPB. One can similarly refer to the OPM $M_{31}$ as the three-qubit OPB consisting of $\ket{0,0,0},\ket{a,1,0},...,\ket{1,1,1}$.

Actually, it has been proven that every two-qubit and three-qubit OPB is equivalent to an OPB of one of the OPMs in \eqref{eq:M2} and \eqref{eq:M32} in \cite{cdjpa2017}. 
The same reference has also classified four-qubit OPBs, while it is not easy to further classify $n$-qubit OPBs with $n>4$. We list the four-qubit OPBs as follows, because we will use them in the next sections.

\begin{lemma}
\label{le:4qubit}
There are totally 33 four-qubit OPBs which are pairwise not equivalent. We list them as follows, namely matrices $M_{41}-M_{433}$.

\bea
\label{eq:prek-1a}
&& 
M_{41}=
\left[ \begin{array}{cccc}
0 & 0 & 0 & * \\
0 & 0 & 1 & * \\
0 & 1 & c & * \\
0 & 1 & c' & * \\
1 & b & x & * \\
1 & b & x' & * \\
1 & b' & y & * \\
1 & b' & y' & * \\  
\end{array} \right],
\quad
M_{42}=
\left[ \begin{array}{cccc}
0 & 0 & * & 0 \\ 0 & 0 & * & 1 \\
0 & 1 & 0 & * \\
0 & 1 & 1 & * \\
1 & b & c & * \\
1 & b & c' & * \\
1 & b' & x & * \\
1 & b' & x' & * \\
\end{array} \right],
\quad
M_{43}=
\left[ \begin{array}{cccc}
0 & 0 & 0 & * \\
0 & 0 & 1 & * \\
0 & 1 & * & 0 \\
0 & 1 & * & 1 \\
1 & b & * & d \\
1 & b & * & d' \\
1 & b' & c & * \\
1 & b' & c' & * \\
\end{array} \right],
\quad
\\  \label{eq:prek-1b}
&& 
M_{44}=\left[ \begin{array}{cccc}
0 & 0 & 0 & * \\
0 & 0 & 1 & * \\
0 & 1 & c & * \\
0 & 1 & c' & * \\
1 & b & * & 0 \\
1 & b & * & 1 \\
1 & b' & * & d \\
1 & b' & * & d' \\
\end{array} \right],
\quad
M_{45}=
\left[\begin{array}{cccc}
0 & 0 & * & 0 \\
0 & 0 & * & 1 \\
0 & 1 & c & * \\
0 & 1 & c' & * \\
1 & b & 0 & * \\
1 & b' & 0 & * \\
1 & x & 1 & * \\
1 & x' & 1 & * \\
\end{array} \right],
\quad
M_{46}=
\left[ \begin{array}{cccc}
0 & 0 & * & 0 \\
0 & 0 & * & 1 \\
0 & 1 & * & d \\
0 & 1 & * & d' \\
1 & b & 0 & * \\
1 & b' & 0 & * \\
1 & x & 1 & * \\
1 & x' & 1 & * \\
\end{array} \right],
\eea

\bea
\label{eq:prek-2}
&& 
M_{47}=\left[ \begin{array}{cccc}0 & 0 & c & * \\
0 & 0 & c' & * \\
0 & 1 & x & * \\
0 & 1 & x' & * \\
1 & b & 0 & 0 \\
1 & * & 1 & 0 \\
1 & b & * & 1 \\
1 & b' & 0 & * \\
1 & b' & 1 & 1 \\
\end{array} \right]
,\quad
M_{48}=
\left[ \begin{array}{cccc}
0 & 0 & * & d \\
0 & 0 & * & d' \\
0 & 1 & c & * \\
0 & 1 & c' & * \\
1 & * & 1 & 0 \\
1 & b & 0 & 0 \\
1 & b & * & 1 \\
1 & b' & 0 & * \\
1 & b' & 1 & 1 \\
\end{array} \right]
,\quad
\eea

\bea
\label{eq:prek-3a}
&&
M_{49}=
 \left[ \begin{array}{cccc}
0 & 0 & 0 & * \\
0 & b & 1 & * \\
0 & b' & 1 & * \\
1 & 0 & c & * \\
1 & 0 & c' & * \\
1 & 1 & 1 & * \\
a & 1 & 0 & * \\
a' & 1 & 0 & * \\
\end{array} \right]
,\quad
M_{410}=
\left[ \begin{array}{cccc}
0 & 0 & 0 & * \\
0 & b & 1 & * \\
0 & b' & 1 & * \\
1 & 0 & * & 0 \\
1 & 0 & * & 1 \\
1 & 1 & 1 & * \\
a & 1 & 0 & * \\
a' & 1 & 0 & * \\
\end{array} \right]
,\quad
\\  \label{eq:prek-3b}
&& 
M_{411}=\left[ \begin{array}{cccc}
0 & 0 & 0 & * \\
0 & * & 1 & 0 \\
0 & * & 1 & 1 \\
1 & 0 & * & d \\
1 & 0 & * & d' \\
1 & 1 & 1 & * \\
a & 1 & 0 & * \\
a' & 1 & 0 & * \\
\end{array} \right]
,\quad
M_{412}=
\left[ \begin{array}{cccc}
0 & 0 & 0 & * \\
0 & * & 1 & 0 \\
0 & * & 1 & 1 \\
1 & 0 & * & d \\
1 & 0 & * & d' \\
1 & 1 & 1 & * \\
* & 1 & 0 & x \\
* & 1 & 0 & x' \\
\end{array} \right]
,\quad
\eea

\bea
\label{eq:prek-5a}
&& 
M_{413}=\left[ \begin{array}{cccc}
0 & 0 & 0 & * \\
0 & 1 & 0 & 1 \\
0 & b & 1 & * \\
0 & b' & 1 & * \\
1 & 1 & 1 & 0 \\
1 & 1 & * & 1 \\
1 & 0 & c & * \\
1 & 0 & c' & * \\
* & 1 & 0 & 0 \\
\end{array} \right]
,\quad
M_{414}=
\left[ \begin{array}{cccc}
0 & 0 & 0 & * \\
0 & 1 & 0 & 1 \\
0 & b & 1 & * \\
0 & b' & 1 & * \\
1 & 1 & 1 & 0 \\
1 & 1 & * & 1 \\
1 & 0 & * & d \\
1 & 0 & * & d' \\
* & 1 & 0 & 0 \\
\end{array} \right]
,\quad
\\  \label{eq:prek-5b}
&& 
M_{415}=\left[ \begin{array}{cccc}
0 & 0 & 0 & * \\
0 & 1 & 0 & 1 \\
0 & * & 1 & d \\
0 & * & 1 & d' \\
1 & 1 & 1 & 0 \\
1 & 1 & * & 1 \\
1 & 0 & c & * \\
1 & 0 & c' & * \\
* & 1 & 0 & 0 \\
\end{array} \right]
,\quad
M_{416}=
\left[ \begin{array}{cccc}
0 & 0 & 0 & * \\
0 & 1 & 0 & 1 \\
0 & * & 1 & d \\
0 & * & 1 & d' \\
1 & 1 & 1 & 0 \\
1 & 1 & * & 1 \\
1 & 0 & * & x \\
1 & 0 & * & x' \\
* & 1 & 0 & 0 \\
\end{array} \right]
,\quad
\eea

\bea
\label{eq:prek-6}
&& 
M_{417}=\left[ \begin{array}{cccc}
0 & 0 & 0 & 0 \\
0 & * & 0 & 1 \\
0 & b & 1 & * \\
0 & b' & 1 & * \\
1 & 1 & 1 & 0 \\
1 & 1 & * & 1 \\
1 & 0 & c & * \\
1 & 0 & c' & * \\
* & 1 & 0 & 0 \\
\end{array} \right]
,\quad
M_{418}=
\left[ \begin{array}{cccc}
0 & 0 & 0 & 0 \\
0 & 0 & * & 1 \\
0 & 1 & * & d \\
0 & 1 & * & d' \\
1 & 1 & 1 & 0 \\
1 & * & 1 & 1 \\
1 & b & 0 & * \\
1 & b' & 0 & * \\
* & 0 & 1 & 0 \\
\end{array} \right]
,\quad
M_{419}=
\left[ \begin{array}{cccc}
0 & 0 & 0 & 0 \\
0 & * & 0 & 1 \\
0 & * & 1 & d \\
0 & * & 1 & d' \\
1 & 1 & 1 & 0 \\
1 & 1 & * & 1 \\
1 & 0 & * & x \\
1 & 0 & * & x' \\
* & 1 & 0 & 0 \\
\end{array} \right]
,\quad
\eea

\bea
\label{eq:prek-7}
&& 
M_{420}=\left[ \begin{array}{cccc}
0 & 0 & 0 & 0 \\
0 & 0 & 1 & * \\
0 & 1 & c & * \\
0 & 1 & c' & * \\
1 & 1 & 1 & 0 \\
1 & 1 & 0 & * \\
1 & 0 & * & 0 \\
1 & * & 1 & 1 \\
* & 0 & 0 & 1 \\
\end{array} \right]
,\quad
M_{421}=
\left[ \begin{array}{cccc}
0 & 0 & 0 & 0 \\
0 & 0 & 1 & * \\
0 & 1 & * & d \\
0 & 1 & * & d' \\
1 & 1 & 1 & 0 \\
1 & 1 & 0 & * \\
1 & 0 & * & 0 \\
1 & * & 1 & 1 \\
* & 0 & 0 & 1 \\
\end{array} \right]
,\quad
\eea

\bea
\label{eq:prek-8}
&& 
M_{422}=\left[ \begin{array}{cccc}
0 & 0 & 0 & 0 \\
0 & 1 & 0 & d \\
0 & * & 1 & d \\
0 & 1 & * & d' \\
0 & 0 & 1 & d' \\
1 & 1 & c & * \\
1 & 1 & c' & * \\
1 & 0 & 1 & 1 \\
1 & 0 & * & 0 \\
* & 0 & 0 & 1 \\
\end{array} \right]
,\quad
M_{423}=
\left[ \begin{array}{cccc}
0 & 0 & 0 & 0 \\
0 & 1 & 0 & d \\
0 & * & 1 & d \\
0 & 1 & * & d' \\
0 & 0 & 1 & d' \\
1 & 1 & * & x \\
1 & 1 & * & x' \\
1 & 0 & 1 & 1 \\
1 & 0 & * & 0 \\
* & 0 & 0 & 1 \\
\end{array} \right]
,\quad
\eea

\bea
\label{eq:prek-9}
&& 
M_{424}=\left[ \begin{array}{cccc}
0 & 0 & 0 & 0 \\
0 & b & 0 & 1 \\
0 & b & 1 & * \\
0 & b' & 1 & 0 \\
0 & b' & * & 1 \\
1 & 1 & 0 & 1 \\
1 & 1 & 1 & * \\
1 & 0 & d & * \\
1 & 0 & d' & * \\
* & 1 & 0 & 0 \\
\end{array} \right]
,\quad
M_{425}=
\left[ \begin{array}{cccc}
0 & 0 & 0 & 0 \\
0 & b & 0 & 1 \\
0 & b & 1 & * \\
0 & b' & 1 & 0 \\
0 & b' & * & 1 \\
1 & 1 & 0 & 1 \\
1 & 1 & 1 & * \\
1 & 0 & * & d \\
1 & 0 & * & d' \\
* & 1 & 0 & 0 \\
\end{array} \right]
,\quad
\eea

\bea
\label{eq:prek-10}
&& 
M_{426}=\left[ \begin{array}{cccc}
0 & 0 & 0 & 0 \\
0 & 1 & 1 & 1 \\
0 & * & 1 & 0 \\
0 & 0 & * & 1 \\
1 & 1 & 1 & d \\
1 & 0 & 0 & d' \\
1 & * & 1 & d' \\
1 & 0 & * & d \\
a & 1 & 0 & * \\
a' & 1 & 0 & * \\
\end{array} \right]
,\quad
M_{427}=
\left[ \begin{array}{cccc}
0 & 0 & 0 & 0 \\
0 & 1 & 1 & 1 \\
0 & * & 1 & 0 \\
0 & 0 & * & 1 \\
1 & 1 & 1 & d \\
1 & 0 & 0 & d' \\
1 & * & 1 & d' \\
1 & 0 & * & d \\
* & 1 & 0 & x \\
* & 1 & 0 & x' \\
\end{array} \right] 
,
\eea

\bea \label{eq:sw10-12}
&& 
M_{428}=\left[ \begin{array}{cccc}
0 & 0 & 0 & 0 \\
0 & * & 1 & 0 \\
0 & 0 & * & 1 \\
0 & 1 & 0 & * \\
0 & 1 & 1 & 1 \\
1 & b & c & d \\
1 & * & c' & d \\
1 & b & * & d' \\
1 & b' & c & * \\
1 & b' & c' & d' \\
\end{array} \right],
\quad
M_{429}=
\left[ \begin{array}{cccc}
0 & * & 1 & 0 \\
0 & 0 & * & 1 \\
0 & 1 & 0 & * \\
1 & * & 0 & 1 \\
1 & 1 & * & 0 \\
1 & 0 & 1 & * \\
* & 0 & 0 & 0 \\
* & 1 & 1 & 1 \\
\end{array} \right], 
\quad
M_{430}=
\left[ \begin{array}{cccc}
0 & 0 & 0 & 0 \\
0 & * & 1 & 0 \\
0 & 0 & * & 1 \\
0 & 1 & 0 & * \\
1 & 1 & 1 & 0 \\
1 & 0 & 1 & d \\
1 & * & 0 & d \\
1 & 0 & * & d' \\
1 & 1 & 0 & d' \\
* & 1 & 1 & 1 \\
\end{array} \right],
\end{eqnarray}


\bea \label{eq:sw13-15}
&& 
M_{431}=\left[ \begin{array}{cccc}
0 & 0 & 0 & 0 \\
0 & 1 & 0 & d \\
0 & * & 1 & d \\
0 & 1 & * & d' \\
0 & 0 & 1 & d' \\
1 & 0 & 1  & 1\\
1 & 0 & c & 0\\
1 & 1 & c & *\\
1 & * & c' & 0 \\
1 & 1 & c' & 1 \\
* & 0 & 0 & 1 \\
\end{array} \right],
\quad
M_{432}=
\left[ \begin{array}{cccc}
0 & 0 & 0 & 0 \\
0 & 1 & 1 & d \\
0 & 0 & * & 1 \\
0 & 1 & * & d' \\
1 & 1 & 1 & 0 \\
1 & 0 & 0 & d \\
1 & * & 0 & d' \\
1 & * & 1 & 1 \\
* & 0 & 1 & 0 \\
* & 1 & 0 & d \\
\end{array} \right],
\quad
M_{433}=
\left[ \begin{array}{cccc}
0 & 0 & 0 & 0 \\
0 & 0 & c & 1 \\
0 & 1 & 0 & d \\
0 & * & 1 & 0 \\
0 & 1 & 1 & 1 \\
1 & 0 & c' & 0 \\
1 & 0 & c & d' \\
1 & 1 & c' & d \\
1 & * & c & d \\
1 & 1 & 1 & d' \\
* & 0 & c' & 1 \\
* & 1 & 0 & d' \\
\end{array} \right].
\end{eqnarray}
\qed
\end{lemma}

To conclude this section, we shall refer to an $n$-partite OPB as an \textit{$A_1$-reducible} OPB when up to equivalence, the OPB consists of $\ket{a_1}\otimes\cT_1,...,\ket{a_k}\otimes\cT_k$ and $\ket{a_{k+1}}\otimes\cT_{k+1},...,\ket{a_{d_1}}\otimes\cT_{d_1}$, such that there is an integer $k$, and $\ket{a_i}\perp\ket{a_j}$ for any $i\in[1,k]$ and any $j\in[k+1,d_1]$, and $\cT_j$ is a set of $(n-1)$-partite orthogonal product states for every $j$. We shall say that $\cS_n$ is reducible when it is $A_j$-reducible for some $j$. If $\cS_n$ is not reducible, then we say that $\cS_n$ is irreducible. Then the following observation is clear.
\begin{lemma}
\label{le:reducible=SUMirreducible}	
Every reducible OPB is the union of a few irreducible OPBs. Each irreducible OPB can be obtained by a local projection on the reducible OPB.
\end{lemma}
By definition, one can see that the irreducible OPBs are pairwise orthogonal. Each irreducible OPB span a subspace $\bbC^{f_1}\otimes...\otimes\bbC^{f_n}$ and $1\le f_j\le d_j$.

Next we present the following fact. 


\begin{lemma}
\label{le:multiplicity<n}
Suppose $\cS_n=\{\ket{a_{j,1},...,a_{j,n}},j=0,1,...,d_1...d_n-1\}$ is an irreducible $n$-partite OPB in $\bbC^{d_1}\otimes...\otimes\bbC^{d_n}$. Then 

(i) every state $\ket{a_{j,1}}$ in the set $\{\ket{a_{0,1}},...,\ket{a_{d_1...d_n-1,1}}\}$ has multiplicity at most $d_2...d_n-1$;

(ii) Counting multiplicity, a nonzero vector in $\bbC^{d_1}$ is orthogonal to at most $d_1d_2...d_n-d_2...d_n-1$ states in the set $\{\ket{a_{0,1}},...,\ket{a_{d_1...d_n-1,1}}\}$. 	
\end{lemma}
\begin{proof}
(i) Because every two states in $\cS_n$ are orthogonal, the multiplicity of $\ket{a_{j,1}}$ is at most $d_2...d_n$. Suppose some $\ket{a_{j,1}}$ has multiplicity exactly $d_2...d_n$. Because every two states in $\cS_n$ are orthogonal, we obtain that $\ket{a_{j,1}}$ is orthogonal to the remaining states in the set $\{\ket{a_{0,1}},...,\ket{a_{d_1...d_n-1,1}}\}$. So $\cS_n$ is $A_1$ reducible. It is a contradiction with the fact that $\cS_n$ is irreducible. Hence, $\ket{a_{j,1}}$ has multiplicity at most $d_2...d_n-1$.	 

(ii) Let $\ket{\a}$ be a nonzero vector in $\bbC^{d_1}$ and $N=d_1d_2...d_n-d_2...d_n-1$. Suppose $\ket{\a}$ is orthogonal to $N+1$ states in the set $\{\ket{a_{0,1}},...,\ket{a_{d_1...d_n-1,1}}\}$. Up to a subscript permutation, we may assume that they are $\ket{a_{0,1}},...,\ket{a_{N,1}}$. Because $\{\ket{a_{j,1},...,a_{j,n}},j=0,1,...,N\}$ are orthogonal product states, they span a subspace $\cK \otimes\bbC^{d_2}\otimes...\otimes\bbC^{d_n}$, where $\dim\cK=d_1-1$. Because the subspace is orthogonal to $\{\ket{a_{j,1},...,a_{j,n}},j=N+1,N+2,...,N+d_2...d_n\}$, we obtain that $\ket{a_{j,1}}\perp\cK$. Hence, $\cS_n$ is $A_1$-reducible. It is a contradiction with the hypothesis that $\cS_n$ is irreducible. Hence, $\ket{\a}$ is orthogonal to at most $N$ states in the set $\{\ket{a_{0,1}},...,\ket{a_{d_1...d_n-1,1}}\}$. 
\end{proof}

Based on above-mentioned facts, in the next two sections we locally distinguish some OPBs including two, three and four-qubit OPBs. We shall also show the indistinguishability of more OPBs without entanglement.

\section{Local distinguishability of two-qubit, and some three and four-qubit OPBs}
\label{sec:23qubit}

In this section, we shall not spend quantum entanglement in any discrimination protocols. We show that every two-qubit OPB is locally distinguishable in Theorem \ref{thm:2qubit}. We further investigate the distinguishability of three-qubit OPBs in Theorem \ref{thm:3qubit}. We also investigate the  local  distinguishability of four-qubit OPBs in Theorem \ref{thm:4qubit}.

\begin{theorem} 
\label{thm:2qubit}
Every two-qubit OPB is locally distinguishable. 
\end{theorem}
\begin{proof}
Using Lemma \ref{le:OPB=distinguishableATtheSAMEtime}, it suffices to distinguish the OPB with first OPM in \eqref{eq:M2}. As the first step, Alice performs the POVM $\{\proj{0},\proj{1}\}$ on her particle. If the measurement result is $\proj{0}$, then she informs Bob of the result, so that Bob measures his particle by the POVM $\{\proj{b},\proj{b'}\}$. Similarly, if the measurement result is $\proj{1}$, then Alice informs Bob of the result, so that Bob measures his particle by the POVM $\{\proj{c},\proj{c'}\}$. In both case, Bob can obtain the final result, and thus finishes the discrimination task. 
\end{proof}

The result is heavily based on the simple classification of two-qubit OPBs. The case becomes more complex for three-qubit OPBs, as we show below.

\begin{theorem}
\label{thm:3qubit}
(i) The three-qubit OPB equivalent to $M_{31}$ in \eqref{eq:M2} is locally indistinguishable when none of the vector variables $a,b,c$ are in $\{0,1\}$.

(ii) The three-qubit OPB equivalent to $M_{31}$ in \eqref{eq:M2} is locally distinguishable when one of $a,b,c$ is in $\{0,1\}$.

(iii) The three-qubit OPBs equivalent to $M_{32}$ and $M_{33}$ in \eqref{eq:M32} are locally distinguishable. 
\end{theorem}
\begin{proof}
Using Lemma \ref{le:OPB=distinguishableATtheSAMEtime}, it suffices to prove the assertion for $M_{31},M_{32}$ and $M_{33}$ in \eqref{eq:M2} and \eqref{eq:M32}. 

(i), (ii) First of all, we show that $M_{31}$ in \eqref{eq:M2} is locally indistinguishable when none of the vector variables $a,b,c$ are in $\{0,1\}$. One can show that, if we switch the columns of the second OPM in \eqref{eq:M2}, then by switching $\ket{0}$ and $\ket{1}$ by a local unitary gate $\s_x$, the OPM is unchanged. Hence, it suffices to show that, $M_{31}$ cannot be locally distinguished by starting with Alice. 
Suppose Alice performs the POVM $\{A_j^\dg A_j\}$ on $M_{31}$, such that $\sum_j A_j^\dg A_j=I$. Then the resulting states become $A_j\ket{0}\otimes\ket{0,0}, A_j\ket{a}\otimes\ket{1,0}, A_j\ket{a'}\otimes\ket{1,0}, A_j\ket{0}\otimes\ket{b,1}, A_j\ket{0}\otimes\ket{b',1}, A_j\ket{1}\otimes\ket{0,c}, A_j\ket{1}\otimes\ket{0,c'}, A_j\ket{1}\otimes\ket{1,1}.$ To make the resulting states of system $B,C$ distinguishable, they should be pairwise orthogonal. Hence the order-two positive semidefinite matrix $A_j^\dg A_j$ is diagonal. 
Next, the orthogonality between 
$A_j\ket{a}\otimes\ket{1,0}$ and $ A_j\ket{a'}\otimes\ket{1,0}$ implies that $A_j^\dg A_j$ is a scalar matrix. Hence, it is impossible to distinguish $M_{31}$ under LOCC.

On the other hand, if one of $a,b,c$ is in $\{0,1\}$, then $M_{31}$ evidently becomes $M_{32}$ or $M_{33}$. We shall show that they are both locally distinguishable below.

(iii) Next we distinguish the three-qubit OPB $M_{32}$. Suppose Alice performs the POVM $\{\proj{0},\proj{1}\}$. If the measurement result is $\proj{0}$ then \eqref{eq:M32} implies that system $B,C$ are in a state of the two-qubit OPB $\ket{0,a},\ket{0,a'},\ket{1,b},\ket{1,b'}$. It is distinguishable in terms of Lemma \ref{le:OPB=distinguishableATtheSAMEtime}. 
On the other hand if the measurement result is $\proj{1}$, then \eqref{eq:M32} implies that system $B,C$ are in a state of the two-qubit OPB $\ket{e,x},\ket{e,x'},\ket{e',y},\ket{e',y'}$. It is distinguishable in terms of Lemma \ref{le:OPB=distinguishableATtheSAMEtime}. We have shown that the three-qubit OPB $M_{32}$ in \eqref{eq:M32} is distinguishable. One can similarly show that the three-qubit OPB $M_{33}$ in \eqref{eq:M32} is distinguishable. We have proven the assertion.
\end{proof}

The fact shows that due to the increasing complexity of three-qubit system, one cannot locally distinguish all three-qubit OPBs. So the nonlocality without entanglement can be manifested by the three-qubit OPBs in \eqref{eq:M2}. This observation can also be extended to four-qubit OPBs in \eqref{eq:prek-1a}-\eqref{eq:sw13-15} by a more involved argument, as we show below. 

\begin{theorem}
\label{thm:4qubit}	
For the four-qubit OPBs $M_{41},M_{42},...,M_{433}$ in \eqref{eq:prek-1a}-\eqref{eq:sw13-15}, 

(i) the OPBs $M_{41},M_{42},...,M_{46}$ in \eqref{eq:prek-1a}-\eqref{eq:prek-1b} are locally distinguishable;

(ii) the OPBs $M_{47},M_{48},...,M_{433}$ in \eqref{eq:prek-2}-\eqref{eq:sw13-15} are locally distinguishable when up to equivalence, each of them can be written as $\{\ket{0}\otimes M_0,\ket{1}\otimes M_1\}$ with the three-qubit OPBs $M_0,M_1$ equivalent to $M_{32}$ or $M_{33}$ in \eqref{eq:M32};

(iii) the OPBs $M_{47},M_{48},...,M_{433}$ in \eqref{eq:prek-2}-\eqref{eq:sw13-15} are locally indistinguishable when up to equivalence, none of them can be written as $\{\ket{0}\otimes M_0,\ket{1}\otimes M_1\}$ with the three-qubit OPBs $M_0,M_1$ equivalent to $M_{32}$ or $M_{33}$ in \eqref{eq:M32}.
\end{theorem}
\begin{proof}
(i) One can verify that, each of the six OPBs $M_{41},M_{42},...,M_{46}$ can be written as $\{\ket{0}\otimes M_0,\ket{1}\otimes M_1\}$ with $M_0,M_1$ equivalent to $M_{32}$ or $M_{33}$ in \eqref{eq:M32}. They are locally distinguishable by Theorem \ref{thm:3qubit} (iii).

(ii) When the system $A_1$ performs the POVM $\{\proj{0},\proj{1}\}$, 
the remaining three systems may be in a state from the OPB $M_0$ or $M_1$. 
It follows from Theorem \ref{thm:3qubit} that $M_0$ and $M_1$ are both locally distinguishable. So the assertion holds. 

(iii) Let $M$ be one of the OPBs $M_{47},M_{48},...,M_{433}$. We use the POVM $\{N_j^\dg N_j\}$ such that $N_j=A_j\otimes B_j\otimes C_j \otimes D_j$ for every $j$. Note that $A_j$ is non-unitary, otherwise it is meaningless to perform $A_j$. If $A_j$ does not kill $\ket{0}$ and $\ket{1}$, then one can see that any one of the resulting set has more than eight product states. So it is impossible for Bob, Charlie and Daniel to distinguish the resulting set. On the other hand, if $A_j$ kills $\ket{0}$ or $\ket{1}$, then the resulting set also has more than eight product states. So it is impossible for Bob, Charlie and Daniel to distinguish the resulting set.   
\end{proof}

So far, we have constructed some locally non-distinguishable three and four-qubit OPBs, and thus we have derived the nonlocality from them. They can be distinguished only by using quantum protocols assisted by entanglement, such as teleportation. We shall proceed with this idea in the next section.

\section{Local distinguishability of three and four-qubit OPBs assisted by quantum teleportation}
\label{sec:4qubit}

In this section, we locally distinguish three and four-qubit OPBs using one ebit. It is known that every $2\times d$ OPB is reducible and locally distinguishable  \cite{BDF+99}. By teleportation, one can merge two systems. So we have the following fact.

\begin{theorem}
\label{thm:3qubit-qt}
Every three-qubit OPB in \eqref{eq:M2} is locally distinguishable by using one ebit.	
\end{theorem}	

In the following we distinguish four-qubit OPBs in \eqref{eq:prek-2}-\eqref{eq:sw13-15}
assisted by entanglement. Evidently, two ebits are sufficient, because they can be used to teleport the qubits $C$ and $D$ to system $A$ at the same time. Nevertheless, two ebits may be not necessary for some four-qubit OPBs in \eqref{eq:prek-2}-\eqref{eq:sw13-15}. 
For example, we only need merge system $C,D$ in the four-qubit OPB $M_{47}$ and $M_{48}$ in \eqref{eq:prek-2}, by using one ebit via quantum teleportation. In the following, we investigate the local distinguishability of OPBs in \eqref{eq:prek-3a}-\eqref{eq:sw13-15}.


\begin{theorem}
\label{thm:4qubit-qt}
Every one of the four-qubit OPBs $M_{49},M_{410},M_{411},M_{412}$ in \eqref{eq:prek-3a}-\eqref{eq:prek-3b} is locally distinguishable by merging system $A$ and $B$, using at most one ebit.	
\end{theorem}	
\begin{proof}
We prove the assertion for $M_{49}$ only, and one can similarly prove the assertion for $M_{410},M_{411},M_{412}$. Suppose Alice obtains the qubit of Bob by teleportation. So Alice can measure both of the first two qubits, by using the POVM $\{\proj{10},I-\proj{10}\}$. If the measurement result is $\proj{10}$ then C and D can distinguish the remain state from $\{\ket{c,*},\ket{c',*}\}$. On the other hand, if the measurement result is $I-\proj{10}$ then Charlie measures the remaining 12 states using the POVM $\{\proj{0},\proj{1}\}$. They respectively result in two sets of six orthogonal tripartite product states, which can be locally distinguished. We have finished the proof for the assertion with $M_{49}$. 	
\end{proof}
One can similarly distinguish the remaining OPBs in \eqref{eq:prek-5a}-\eqref{eq:sw13-15}, though a more detailed analysis is required and more entanglement cost may be necessary.

%
%
%


\section{CONCLUSIONS}
\label{sec:con}

In this paper, using the existing two, three and four-qubit OPBs, we have classified them into OPBs of local distinguishability and non-distinguishability. We have also managed to distinguish every three and four-qubit OPBs using teleportation assisted by Bell states. Some problems arise from this paper. One need distinguish five-qubit OPBs, though they are not fully classified in literature yet as far as we know. Besides, the connection between some OPBs and quantum security may be further studied based on their local non-distinguishability. Further, evaluating the necessary cost of quantum entanglement in system merge of multiqubit OPBs remains an involved problem. Another open problem is whether every irreducible OPB is locally indistinguishable.

\section*{Acknowledgments}

LC was supported by the NNSF of China (Grant No. 11871089), and the Fundamental Research Funds for the Central Universities (Grant Nos. KG12040501, ZG216S1810 and ZG226S18C1). 

\section*{Conflict of interest statement}

On behalf of all authors, the corresponding author states that there is no conflict of interest. 

\section*{Data availability statement} 

All data, models, and code generated or used during the study appear in the submitted article.

\appendix


\bibliographystyle{unsrt}

\bibliography{channelcontrol}


\end{document}